\documentclass[12pt,a4paper,reqno]{article}

\usepackage{amsmath,latexsym,amssymb,amsthm}
\usepackage{graphicx}
\usepackage{hyperref}
\usepackage{amsthm}

\newcommand{\head}{\text{head}}
\newcommand{\tail}{\text{tail}}
\newcommand{\linespace}{\vspace{\baselineskip}}

\newcommand{\upline}{\vspace{-\baselineskip}}

\newcommand{\be}{\begin{equation}}
\newcommand{\ee}{\end{equation}}

\renewcommand{\d}{\text{d}}
\newcommand{\e}{\mathrm{e}}
\newcommand{\f}{\text{f}}
\newcommand{\h}{\text{h}}
\renewcommand{\i}{\mathrm{i}}
\renewcommand{\t}{\text{t}}

\newcommand{\B}{\text{B}}

\newcommand{\Cop}{\text{C}}
\newcommand{\E}{\text{E}}

\renewcommand{\H}{\mathcal{H}}

\renewcommand{\S}{\mathcal{S}}

\newcommand{\<}{\langle}
\renewcommand{\>}{\rangle}
\newcommand{\half}{\tfrac{1}{2}}

\newcommand{\third}{\tfrac{1}{3}}

\newcommand{\ok}{\text{ok}}
\newcommand{\fail}{\text{fail}}
\newtheorem{theorem}{Theorem}

\theoremstyle{definition}

\theoremstyle{remark}

\numberwithin{equation}{section}
\newcommand{\thmref}[1]{Theorem~\ref{#1}}
\newcommand{\secref}[1]{Section~\ref{#1}}

\begin{document}

\title{\bf{Histories without collapse}}

\author{Anthony Sudbery$^1$\\[10pt] \small Department of Mathematics,
University of York, \\[-2pt] \small Heslington, York, England YO10 5DD\\
\small $^1$ tony.sudbery@york.ac.uk}

\date{}

\maketitle

Keywords: quantum mechanics, foundations, histories

\linespace

\begin{abstract}

This paper is a comparison of two theories of the probability of a history in quantum mechanics. One is derived from Copenhagen quantum mechanics using the projection postulate and is the basis of the ``consistent histories'' interpretation; the other is based on a proposal by Bell, originally for the ``pilot state'' theory but here applied to pure unitary quantum mechanics. The first can be used for a wider class of histories but depends on the projection postulate, or ``collapse'', which is widely held to be an unsatisfactory feature of the theory; the second can be used in a theory of the universal state vector without collapse. We examine a simple model based on Wigner's friend, in which Bell's model and the projection postulate give different probabilities for the histories of a sentient system. We also examine the Frauchiger-Renner extension of this model, in which comparison of the two calculations of histories throws light on the contradiction found by Frauchiger and Renner. By extending the model to equip the observer with a memory, we reduce the probability of histories to the use of the Born rule at a single time, and show that the Born rule, with the memory, gives the same result as applying projection in the course of the history, because of entanglement with the memory. Entanglement implements collapse. We discuss the implications of this for the use of histories in quantum cosmology.  

\end{abstract}

\newpage

{\bf Declarations}

\linespace

Funding: This research was not supported by any funding.

Conflicts of interest: I am not aware of any conflict of interest.

Availabiilty of data and material: There is no data or material to be made available.

Code availability: There is no relevant code.

\newpage

\section{Introduction}

In quantum mechanics one often wants to know the probability of a sequence of measurements, or the conditional probability of a measurement result at one time given a result at an earlier time. In other words, one wants the probability of a history. In the form of quantum mechanics presented in most textbooks (the ``Copenhagen interpretation''), this is managed by assuming that the unitary evolution is interrupted by a projection, or ``collapse'', at the times of the sequence. Other interpretations of quantum mechanics, usually associated with the name of Everett, deny that such collapses take place at a fundamental level, and assert that all evolution follows a continuous unitary process. If this is so, however, it is not obvious how to calculate the probability of a history: the only point at which probability appears to enter the theory is via the Born rule at one time. However, in 1984 John Bell \cite{Bell:beables}, in a study of quantum field theory in the context of Bohmian ``pilot state'' theory, proposed a process which can be generalised, detached from the Bohmian context of ``beables'' (or ``hidden variables''), and applied to any history in the different ontology of Everettian quantum mechanics. In this paper the generalisation of Bell's process is presented and applied to some examples deriving from the scenario of Wigner's friend, which bring out particularly clearly the issues related to collapse. The probabilities derived from Bell's process are compared with those obtained from Copenhagen quantum mechanics. Finally, we consider how the presence of memory makes it possible to reduce the probability of a history to a single-time probability, using the Born rule.

The paper is organised as follows. In \secref{histprob} we set up the general formalism for a discussion of histories. The next three sections set out the two methods (Copenhagen and Bell) for calculating of the probability of a history, with a discussion (\secref{Everett}) of the problems of histories and probability in pure unitary quantum mechanics. In \secref{Bell} the solution of these problems based on Bell's process is introduced and applied to a variant of the Wigner's friend scenario due to Brukner; the results are compared with those of the Copenhagen calculation. In \secref{FR} the two methods are applied to a more elaborate version of the Wigner's friend scenario due to Frauchiger and Renner. \secref{memory} shows the effects of memory in this scenario, and the paper concludes with a discussion of the results. 

\section{Histories and probabilities}\label{histprob}

A \emph{history} is a sequence of events $E_0,\ldots, E_n$ at times $t_0,\ldots, t_n$. In quantum mechanics, as in statistics and law (and unlike the study of human history, relativity theory or high-energy physics) an event is defined by a proposition and therefore, in quantum mechanics, by a closed subspace of Hilbert space, or equivalently by the orthogonal projection onto this subspace. A history is therefore given by a time-labelled sequence 
\be\label{history}
h = \big((\Pi_0, t_0), \ldots , (\Pi_n, t_n)\big).
\ee
of events $(\Pi_j, t_j)$ each defined by a projection $\Pi_j$ and a time $t_j$.
For simplicity, we will assume that the system starts in a pure state $|\Psi_0\>$ at time $t = 0$, so that $\Pi_0 = |\Psi_0\>\<\Psi_0|$ and $t_0 = 0$.  The theory should give a probability for this history, i.e. for the truth of the proposition represented by the succession of projections $\Pi_j$ at the times $t_j$. A closely related quantity is the \emph{conditional} probability of an event at one time conditional on an event at an earlier time. However, there are competing accounts of this probability. We now consider two of these accounts.

\section{The Copenhagen probability}\label{Copenhagen}

In the version of quantum mechanics presented in most textbooks, the probability of a history like $h$ in \eqref{history} is obtained by applying the Born rule to successive measurements of the projectors $\Pi_j$, assuming orthogonal projection of the state after each measurement and unitary evolution according to the Schr\"odinger equation between the measurements. This gives the probability
\begin{align}\label{PCop}
P_\text{C}(h) &= \<\Psi_0|\widetilde{\Pi}_1\ldots\widetilde{\Pi}_n\ldots\widetilde{\Pi}_1|\Psi_0\>\\
&= \<K(h)|K(h)\>
\end{align}
where $\widetilde{\Pi}_j = \e^{\i Ht_j}\Pi_j\e^{-\i Ht_j}$ (taking $\hbar =1$) and 
\be\label{histstate}
|K(h)\> = \widetilde{\Pi}_n\ldots\widetilde{\Pi}_1|\Psi_0\>.
\ee
In the consistent-histories \cite{Griffiths:book} and temporal-logic \cite{logicfuture} interpretations of quantum mechanics the formula 
\eqref{PCop} is extended to include conjunctions and disjunctions of histories (though this is not essential in the latter interpretation \cite{Bellgreatenterprise}). Logically desirable properties of probability then require a 
consistency condition on the set of histories, namely that $|K(h_1)\>$ and $|K(h_2)\>$ are orthogonal if $h_1$ and $h_2$ are distinct histories. However, if this condition is not satisfied the formula \eqref{PCop} for single histories is still meaningful provided that the evolution in each time interval $[t_j, t_{j+1}]$ is purely unitary.

Recent discussions of conditional probabilities and histories (\cite{Dolby, Giovannetti, Baumann:condprob}) have used the Page-Wootters mechanism to deal with time. The resulting calculations of conditional probabilities are either equivalent to assuming collapse at the events of a history, or proceed by examining the contents of a memory at a single time after all the times of the history. The latter procedure, as will be discussed in \secref{memory}, gives the same results as assuming collapse.

\section{Probability in collapse-free quantum mechanics}\label{Everett}

The derivation of $P_\text{C}$ assumes that, in addition to unitary evolution according to the Schr\"odinger equation, the quantum state undergoes projection at the events in the history. Such projection is denied in some influential interpretations of quantum mechanics, in particular the relative-state interpretation of Everett and Wheeler \cite{Everett, Wheeler}. In this interpretation the truth about the world and its development in time is given by a universal, time-dependent state vector $|\Psi(t)\>$, and nothing else. It is possible to use the Born rule to deduce that, given a projector $\Pi$, the probability that the corresponding proposition is true at time $t$ is $|\<\Psi(t)|\Pi|\Psi(t)\>|^2$, though we need to discuss what this probability means. But if projection processes are removed from the time evolution, the theory offers no connection between events at different times. This might appear to suggest that the probability of the history $h$ of \eqref{history} should be obtained by treating the $(\Pi_j, t_j)$ as independent events, so that the probability of the history $h$ is 
\be\label{PEverett}
P_\E = \<\Psi(t_1)|\Pi_1|\Psi(t_1)\>\<\Psi(t_2)|\Pi_2|\Psi(t_2)\>\cdots\<\Psi(t_n)|\Pi_n|\Psi(t_n)\>.
\ee
But this makes little sense. It does not reflect the continuity of experience, implying as it does that in a very short time interval, no matter how short, there is a finite probability, independent of the length of the interval, of the state changing to a very different one; and it gives a non-zero probability for a history (decayed at $t_1$, undecayed at $t_2$) of an unstable particle, where $t_2 > t_1$. These defects are remedied \cite{verdammte} in the proposal of John Bell which is explained in the next section. First, however, we will look more closely at the meaning of probability in the Everett interpretation.

Probability enters the formalism of quantum mechanics only in the Born rule for the results of measurements and the projection postulate which is associated with measurements. This is rejected in Everettian theory (and indeed, according to Bell \cite{Bell:piddling}, any respectable physical theory) in which there are no preferred physical processes which have the special status of ``measurement''. In Everettian quantum mechanics there is only deterministic evolution of the universal state vector, so there is no absolute role for probability. However, the universe includes sentient systems (or ``observers'', or ``organisms'') and the universal state vector must incorporate the experience of such systems, including the changing experience in time. Each possible experience of a sentient system must be defined by a set of values of physical quantities relating to the organism, so the experience corresponds to a set of eigenvalues of these quantities and therefore to a subspace of the sentient organism's state space. The orthogonal projection onto this experience subspace extends to a projection of the universal state space, corresponding to the proposition ``the organism has the given experience''. Now probability applies to the change in time of the organism's experience: while having experience $E_1$ at time $t_1$, the organism wants to know the probability that it will experience $E_2$ at a later time $t_2$, and then $E_3$ at time $t_3$, and so on. Thus probability is not absolute but \emph{relative} to a particular (appropriately constituted) subsystem, to a particular experience of that subsystem, and to a particular time. 

Note that ``relative'' does not mean ``subjective''; this probability is an objective fact about the organism's situation and does not refer to its beliefs, just as the velocity of a particle is relative to a frame of reference but is an objective property of the particle.  

\section{The Bell probability}\label{Bell}

%The appropriate role of probability in a collapse-free formulation of quantum mechanics is not as an absolute property of the universal state, but as relative to a particular sentient subsystem of the universe, and to a particular experience of that subsystem. Such a sentient being will have a set of observables defining definite experiences, whose eigenspaces define a complete set of subspaces, not only of the state space of the sentient being but also of the universal state space. The changing experience of the sentient being is represented by a state vector moving between these subspaces. This movement can be described by adapting a stochastic model originally proposed by Bell in the context of a Bohmian model of quantum field theory.

Let $\Pi_j$ be the projector onto the subspace $\S_j$ of the universal state space describing experience $\eta_j$ of the sentient being, or ``observer''. We will call these subspaces \emph{sensible} subspaces because of their relation to the sense organs of the observer; the same word will be applied to the corresponding subspaces of the observer's state space (a tensor factor of the universal state space). The experience label $j$ is assumed discrete for convenience.  Adopting the observer's perspective, we assume that at each time $t$ the observer has a definite experience $\eta_j$ (this, however, is not true in the external perspective in which the whole truth is the universal state vector $|\Psi(t)\>$, a superposition of different experience states). To describe the perceived change of this experience in time we assume, following Bell \cite{Bell:beables}, that it follows a stochastic process governed by the universal Hamiltonian $H$ together with the universal state vector $|\Psi(t)\>$, as follows. If, at time $t$, the  observer's experience is $\eta_j$, then the probability that at time $t + \delta t$ their experience has changed to $\eta_k$  is $w_{jk}\delta t$ where the transition probability $w_{jk}$ is given by
\upline

\Large
\be\label{PBell}
\text{\normalsize $w_{jk}$} = \begin{cases}  \frac{2\text{Im}[\hbar^{-1}\<\Psi(t)|\Pi_k H\Pi_j|\Psi(t)\>]}{\<\Psi(t)|\Pi_j|\Psi(t)\>}&\text{\normalsize if this is $\ge 0$},\\
                       \text{\normalsize $0$}                                                  &\text{\normalsize if it is negative.}
         \end{cases}
\ee
\normalsize
The probabilities for the observer's changing experience are then compatible with the Born rule, in the following sense.
\begin{theorem}\label{BellBorn}(\cite{Bell:beables}, \cite{QMPN} p.215)

If the experience state moves between the sensible subspaces with transition probabilities given by \eqref{PBell}, then the probability $P_j(t)$ that it lies in the subspace $\S_j$ at time $t$ is given by the Born rule
\[
 P_j(t) = \<\Psi(t)|\Pi_j|\Psi(t)\>
\]
at all positive times $t$, if the probabilities are so given at the initial time $t = 0$.
\end{theorem}

This theorem is the justification for the transition probabilities \eqref{PBell}. Bell's formula is not uniquely determined by the requirement that the theorem should hold: there is a range of possible transition probabilities with the same property \cite{BacciaDickson}. However, Bell's formula is uniquely natural in applications to decay \cite{verdammte} and measurement processes \cite{singleworld, Hollowood:classical}: it ensures that the underlying direction of change in such processes is always forwards, without intermittent reversals (decay products, for example, recombining to reconstitute the unstable decaying state). This means that the theory is not invariant under time reversal. I regard this as a virtue; for those with different predilections, the range of transition probabilities found in \cite{BacciaDickson} include one which is time-symmetric. However, it should be emphasised that in the theory presented here, the probabilities arise only in the description of the subjective experience of a sentient system; the existence of a preferred direction of time for such a system has been explained within a time-symmetric theory \cite{Rovelli:time}.
 
We use these transition probabilities to define the Bell probability $P_\B(h)$, relative to a particular observer, of a history of experiences of that observer. More generally, Bell's process will give probabilities for any set of histories in which the projectors $\Pi_j$ are chosen from a fixed complete set of projectors. There is no need for the histories to satisfy the decoherence condition of the consistent-histories formulation of quantum mechanics.

Note that our ontology is different from Bell's. Bell intended the subspaces $\S_j$ to have an absolute meaning, being the subspaces in which certain preferred observables, called ``beables'', had definite values, and that it was absolutely true that the state of the world was in one of those subspaces. In the theory adopted here, nothing has absolute existence or truth except the universal state vector. The subspaces $\S_j$ are only defined \emph{if} the universal state space admits a tensor product decomposition in which some factors have the structure of sentient systems (whatever that may be); the focus on a particular one of those systems is arbitrary. (This is somewhat like some other recently expressed interpretations, such as those called ``convivial solipsism'' \cite{conviv} and ``minimalism'' \cite{minimalism}.) In Bell's exposition of Bohmian theory these subspaces are called ``beable'' \cite{Bell:beables} or ``viable'' (\cite{QMPN} p. 215). 

\thmref{BellBorn} usually gives the probabilities for the first stage of a history, but often is not helpful in subsequent stages since the condition that the initial probabilities are given by the Born rule is not satisfied. To illustrate this, we now give the calculation of history probabilities in a version of the ``Wigner's friend'' scenario considered by Brukner \cite{Brukner1}. 

In this scenario an experimenter (``Wigner'') observes a closed laboratory containing another experimenter (``Wigner's friend'') who makes measurements on a quantum system $S$. We take $S$ to be a qubit (e.g.\ an electron spin) with orthogonal states $|1\>$ and $|2\>$, and we simplify Wigner and his friend so that each of them has just three orthogonal states: a ready state $|0\>$ and two states registering the result of a measurement: $|+\>$ and $|-\>$ for Wigner, $|1\>$ and $|2\>$ for his friend (let us call her Frieda). The scenario is as follows.

At $t=0$ the qubit $S$ is in a general state $c_1|1\> + c_2|2\>$ and both Wigner and Frieda are in their ready states. The universal state is therefore
\[
|\Psi(0)\> = \big(c_1|1\> + c_2|2\>\big)_S|0\>_F|0\>_W.
\]
Between $t=0$ and $t=1$ Frieda measures $S$ in the basis $\{|1\>_S, |2\>_S\}$, so that $S$ and $F$ become entangled and the universal state at $t=1$ is 
\begin{align*}
|\Psi(1)\> &= \big(c_1|1\>_S|1\>_F + c_2|2\>_S|2\>_F\big)|0\>_W\\
&= \big(c_1|1\>_{SF} + c_2|2\>_{SF}\big)|0\>_W \quad \text{in an obvious notation.}
\end{align*}
Between $t=1$ and $t=2$ Wigner measures the state of Frieda and her laboratory in a basis which includes the states
\[
|\pm\>_{SF} = \frac{1}{\sqrt{2}}\big(|1\>\pm|2\>\big)_{SF},
\]
so that the universal state at $t=2$ is
\[
|\Psi(2)\> = \frac{1}{\sqrt{2}}(c_1 + c_2)|+\>_{SF}|+\>_W + \frac{1}{\sqrt{2}}(c_1 - c_2)|-\>_{SF}|-\>_W.
\]
Thus the sensible state of $F$ is $|0\>$ at $t=0$ and either $|1\>$ or $|2\>$ at both $t=1$ and $t=2$. For $F$, there are four possible histories
\[
(0jk) = \big((\Pi_0,t=0), (\Pi_j, t=1), (\Pi_k, t=2)\big)
\]
with $j, k \in \{1,2\}$. The probabilities of transition from $|0\>$ at $t = 0$ are given by \thmref{BellBorn} since the probabilities at $t = 0$, namely  $P_0 = 1, P_1 = P_2 = 0$, are indeed given by the Born rule; hence the probability that Frieda finds herself in state $|j\>$ at $t = 1$ is $|c_j|^2$. To find the probability that she experiences a change of state between $t = 1$ and $t =2$, however, we cannot use this theorem, since her probability for the state $|j\>$ at $t = 1$ is either $0$ or $1$, which is not the probability given by the Born rule: her experience of her state is at odds with the universal state. To calculate the probability of a change of state we need to know the Hamiltonian governing the change in the universal state in this time interval, when Wigner is measuring Frieda and her laboratory. Let us assume that the measurement process takes place by a simple rotation, so that the state in the time interval $1 < t <2$ is given by
\begin{multline*}
|\Psi(1 + \tau)\> = \frac{1}{\sqrt{2}}(c_1 + c_2)|+\>_{SF}\big(\cos\half\pi\tau|0\>_W + \sin\half\pi\tau|+\>_W \big)\\
 + \frac{1}{\sqrt{2}}(c_1 - c_2)|-\>_{SF}\big(\cos\half\pi\tau|0\>_W + \sin\half\pi\tau|-\>_W \big)
\end{multline*}
and the components in Frieda's sensible subspaces are
\begin{align*}
\Pi_1|\Psi(1 + \tau)\> &= |1\>_{SF}\Big(c_1\cos\half\pi\tau|0\>_W  + \sin\half\pi\tau\big[\half(c_1 + c_2)|+\> + \half(c_1 - c_2)|-\>\big]_W \Big),\\
\Pi_2|\Psi(1 + \tau)\> &= |2\>_{SF}\Big(c_2\cos\half\pi\tau|0\>_W  + \sin\half\pi\tau\big[\half(c_1 + c_2)|+\> - \half(c_1 - c_2)|-\>\big]_W \Big).
\end{align*}
The Hamiltonian driving this evolution is, taking $\hbar = 1$,
\be\label{HBruk}
H = \frac{\i\pi}{2}\sum_{z = \pm}|z\>\<z|_{SF}\otimes\big[|z\>\<0| - |0\>\<z|\big]_W.
\ee
The matrix element relevant to the transition $|1\>_F \rightarrow |2\>_F$ is
\be\label{matel}
\<\Psi(1+\tau)|\Pi_2H\Pi_1|\Psi(1 +\tau)\> = \frac{\i\pi}{2}\big(|c_1|^2 - |c_2|^2\big)\cos\half\pi\tau\sin\half\pi\tau.
\ee
Suppose $|c_1|^2 \ge |c_2|^2$; then the transition rate for $|1\>_F \rightarrow |2\>_F$ at time $1 + \tau$ is, according to \eqref{PBell},
\begin{align*}
w_{12}(1 +\tau) &= \frac{2\text{Re}[\<\Psi(1 +\tau)|\Pi_2 H\Pi_1|\Psi(1 +\tau)\>]}{\<\Psi(t)|\Pi_1|\Psi(t)\>}\\
&= \frac{\pi}{2}\big(|c_1|^2 - |c_2|^2\big)\frac{\cos\half\pi\tau\sin\half\pi\tau}{|c_1|^2 +\big(|c_1|^2 - |c_2|^2\big)\sin^2\half\pi\tau},
\end{align*}
while the rate for the opposite transition is $w_{21} = 0$. Hence the probabilities $P_1(1 +\tau), P_2(1 +\tau)$ for Frieda's experience at time $1 +\tau$ satisfy
\[
\frac{\d}{\d\tau}P_1 = -w_{12} P_1, \quad \frac{\d}{\d\tau}P_2 = w_{12}P_1,
\]
the solution of which is
\[
P_1(1 + \tau) = C\Big[|c_1|^2 -\half\big(|c_1|^2 - |c_2|^2\big)\sin^2\half\pi\tau\big],\quad P_2 = 1 - P_1
\]
where $C$ is a constant of integration. By taking $C = 1/|c_1|^2$ or $C = 0$, corresponding to initial conditions $P_1(1) = 1$ and $P_1(1) = 0$ respectively, we get the conditional probabilities for Frieda's experience:
\begin{align}\label{Fcond12}
P\big(|1\>_F \text{ at } t = 2\;\big|\;|1\>_F \text{ at } t = 1) &= \frac{1}{2|c_1|^2},\\
P\big(|1\>_F \text{ at } t = 2\;\big|\;|2\>_F \text{ at } t = 1) &= 0,\\
P\big(|2\>_F \text{ at } t = 2\;\big|\;|1\>_F \text{ at } t = 1) &= \frac{|c_1|^2 - |c_2|^2}{2|c_1|^2},\\
P\big(|2\>_F \text{ at } t = 2\;\big|\;|2\>_F \text{ at } t = 1) &= 1.
\end{align}
These, together with the probabilities $|c_1|^2$ and $|c_2|^2$ for the results of Frieda's measurement at $t = 1$, give the non-zero Bell probabilities for her histories as set out in Table 1, which also shows the corresponding Copenhagen probabilities.
\begin{table}[ht]
\begin{center}
\renewcommand{\arraystretch}{1.4}
\begin{tabular}{|c|c|c|}
\hline
$h$ & $P_B(h)$ & $P_C(h)$ \\
\hline\hline
(011) & $\half$ & $\half|c_1|^2$\\
\hline
(012) & $\half\big(|c_1|^2 - |c_2|^2)$ & $\half|c_1|^2$ \\
\hline
(021) & $0$ & $\half|c_2|^2$ \\
\hline
(022)& $|c_2|^2$ & $\half|c_2|^2$ \\
\hline
\end{tabular}
\end{center}
\caption{Probabilities for the histories of Wigner's friend in Brukner's scenario}
\end{table}

Bell and Copenhagen agree that the probability that Frieda is in state $|1\>$ at $t = 1$ is $|c_1|^2$, while the probability that she is in $|2\>$ is $|c_2|^2$, so she is more likely to be in the state $|1\>$, while at $t = 2$ she is equally likely to be in either state. According to Bell, this change is caused entirely by a transition from $|1\>$ to $|2\>$, so she will only feel the effect of Wigner's measurement if she is in the more probable state $|1\>$; but according to Copenhagen there is a flow of probability in both directions.

It is illuminating to examine the two sets of probabilities in the special case $c_1 = c_2 = 1/\sqrt{2}$. for which Bell and Copenhagen agree that the two states $|1\>_F$ and $|2\>_F$ each have probability $\half$ at both $t = 1$ and $t = 2$. According to Bell there are no transitions between these two states in the time period $1 \le t \le 2$, whereas Copenhagen gives equiprobable transitions in both directions. For these values of the coefficients Frieda and her laboratory are already in the state $|+\>_{SF}$ at $t = 1$, so Wigner's measurement simply registers this fact, changing Wigner's state from $|0\>_W$ to $|+\>_W$ but leaving Frieda's state unchanged. This is reflected in the fact that the Hamiltonian \eqref{HBruk} has zero matrix elements \eqref{matel} between the two components $\Pi_i|\Psi(1 +\tau)\>$ in which Frieda has a definite value $i = 1$ or $2$. The Copenhagen calculation of probabilities, however, assumes that Frieda's state collapses at $t = 1$, despite the lack of physical mechanism for this collapse; if this happens then Wigner's measurement does have an effect on the resulting states, leading to transitions between them.

In the general case the Bell calculation gives the minimal transitions necessary for the change in probabilities from $t = 1$ to $t = 2$.

\section{The Frauchiger-Renner universe}\label{FR}

A more elaborate variant of the Wigner's friend thought experiment has been devised by Frauchiger and Renner \cite{FR2}. It is instructive to compare the different probabilities for this scenario. This takes place in a universe consisting of six subsystems $C, S, F_1, F_2, W_1$ and $W_2$.  Of these $C$ is a quantum coin with two orthogonal states $|\head\>_C$ and $|\tail\>_C$; $S$ is a qubit with orthogonal states $|\uparrow\>_S$ and  $|\downarrow\>_S$; and the four systems $F_1, F_2, W_1, W_2$ are sentient agents, each agent $X$ having three orthogonal states $|0\>_X, |a\>_X, |b\>_X$ where $|0\>$ is a state of readiness before $X$ makes a measurement and $|a\>$ and $|b\>$ are states recording the results $a, b$ of the measurement. We refer to the agents $F_1$ and $F_2$ as ``experimenters'', since they perform operations on the physical systems $C$ and $S$, and to $W_1$ and $W_2$ as ``para-experimenters'' since they operate on the experimenters $F_1$ and $F_2$ as well as $C$ and $S$. The experiment proceeds as follows.

Initially, at time $t = 0$, the system is in the state 
\be\label{Psi0}
|\Psi(0)\> = \left(\sqrt{\third}|\head\> + \sqrt{\tfrac{2}{3}}|\tail\>\right)_C|0\>_S|0\>_{F_1}|0\>_{F_2}|0\>_{W_1}|0\>_{W_2}. 
\ee
Between times $t=0$ and $t=1$ the experimenter $F_1$ measures the state of the coin, so that at time $t = 1$ the universal state is 
\be\label{Psi1}
|\Psi(1)\> = \left(\sqrt{\third}|\head\> + \sqrt{\tfrac{2}{3}}|\tail\>\right)_{F_1C}|0\>_S|0\>_{F_2}|0\>_{W_1}|0\>_{W_2},
\ee
where $|\head\>_{F_1C} = |\head\>_{F_1}|\head\>_C$ and similarly for ``tail". 

Between $t = 1$ and $t = 2$ the experimenter $F_1$ prepares the qubit $S$ in a way which depends on the result of their measurement of $C$: if the coin came down showing ``head'', $F_1$ prepares $|\downarrow\>_S$; if it showed ``tail'', $F_1$ prepares $|\rightarrow\>_S = \tfrac{1}{\sqrt{2}}\big(|\uparrow\> + |\downarrow\>\big)_S$. The universal state at $t = 2$ is therefore
\be\label{Psi2}
|\Psi(2)\> = \left(\sqrt{\third}|\head\>_{F_1C}|\downarrow\>_S + \sqrt{\tfrac{2}{3}}|\tail\>_{F_1C}|\rightarrow\>_S\right)|0\>_{F_2}|0\>_{W_1}|0\>_{W_2}.
\ee

Between $t = 2$ and $t = 3$ the qubit $S$ is passed to the second experimenter $F_2$, who measures it in the basis $\{|\uparrow\>, |\downarrow\>\}$, recording the result $+, -$ respectively; we will use the same symbol $\pm$ to denote the joint state of $F_2$ and $S$. The universal state becomes
\be\label{Psi3}
|\Psi(3)\> = \sqrt{\tfrac{1}{3}}\Big(|\head\>_{F_1C}|-\>_{F_2S} + |\tail\>_{F_1C}|+\>_{F_2S} + 
|\tail\>_{F_1C}|-\>_{F_2S}\Big)|0\>_{W_1}|0\>_{W_2}.
\ee

Between $t = 3$ and $t = 4$ the first para-experimenter $W_1$ measures the state of $F_1$ and the coin $C$ in the basis
\begin{align*}
|\ok\>_{F_1C} &= \tfrac{1}{\sqrt{2}}\big(|\head\>_{F_1C} - |\tail\>_{F_1C}\big),\\
|\fail\>_{F_1C} &= \tfrac{1}{\sqrt{2}}\big(|\head\>_{F_1C} + |\tail\>_{F_1}C\big),
\end{align*}
recording the result in their own state $|\ok\>_{W_1}$ or $|\fail\>_{W_1}$. The universal state becomes, at $t = 4$,
\begin{multline*}%\label{Psi4}
|\Psi(4)\> = \Big(\sqrt{\tfrac{1}{6}}\big[ - |\ok\>_{F_1C}|\ok\>_{W_1} + |\fail\>_{F_1C}|\fail\>_{W_1}\big]|+\>_{F_2S}\\
+ \sqrt{\tfrac{2}{3}}|\fail\>_{F_1C}|\fail\>_{W_1}|-\>_{F_2S}\Big)|0\>_{W_2}.
\end{multline*}

Between $t = 4$ and $t = 5$ the second para-experimenter $W_2$ measures the state of $F_2$ and the qubit $S$ in the basis
\begin{align*}
|\ok\>_{F_2S} &= \tfrac{1}{\sqrt{2}}\big(|-\>_{F_2S} - |+\>_{F_2S}\big),\\
|\fail\>_{F_2S} &= \tfrac{1}{\sqrt{2}}\big(|-\>_{F_2S} + |+\>_{F_2S}\big),
\end{align*}
recording the result in their own state $|\ok\>_{W_2}$ or $|\fail\>_{W_2}$. The universal state  becomes, at $t = 5$,
\begin{multline*}\label{Psi5}
|\Psi(5)\> = \sqrt{\tfrac{1}{12}}\Big(|\ok\>_{F_1C}|\ok\>_{W_1} + |\fail\>_{F_1C}|\fail\>_{W_1}\Big)|\ok\>_{F_2S}|\ok\>_{W_2}\\
+ \sqrt{\tfrac{1}{12}}\Big(- |\ok\>_{F_1C}|\ok\>_{W_1} + 3|\fail\>_{F_1C}|\fail\>_{W_1}\Big)|\fail\>_{F_2S}|\fail\>_{W_2}.
\end{multline*}

Now consider the history of this experiment as seen by the experimenter $F_2$ from $t = 1$ to $t = 5$. The possible histories are 
\[
(abcde) = h(a,b,c,d,e) = \big( (\Pi_a, t = 1),\ldots, (\Pi_e, t = 5) \big)
\]
where each of $a,b,c,d,e$ is taken from the set $\{0, +, -\}$, and $\Pi_x$ is the projector onto the 32-dimensional subspace 
\[ 
\S_x = \H_C\otimes \H_S\otimes\H_{F_1}\otimes\text{span}\left(|x\>_{F_2}\right)\otimes\H_{W_1}\otimes\H_{W_2}.
\]
The Copenhagen probabilities $P_\Cop(abcde)$ can be calculated in the same way as in Section \ref{Copenhagen}. The result is that the only non-zero probabilities are 
\[
P_\Cop(00+++) = \tfrac{1}{6}, \quad P_\Cop(00--+) = \tfrac{1}{3}, \quad P_\Cop(00++-) = \tfrac{1}{6}, \quad P_\Cop(00---) = \tfrac{1}{3}.
\]

To calculate the Bell probabilities, we need to know the Hamiltonian governing the evolution of the system, i.e. the succession of measurements made by the four agents.  In each time interval $[n,n+1]$ ($n = 1,\ldots,4$) the system evolves from a product state to an entangled state of the experimenter and the measured system. The Bell probabilities can be calculated, as in \secref{Bell}, by assuming that the Hamiltonian in this interval is a generator of rotations in the subspace spanned by these two states, operating for a time such that $\e^{-\i H}$ is a rotation through $\pi/2$. The resulting non-zero probabilities are 
\[
P_\B(00+++) = \tfrac{1}{3}, \quad P_\B(00--+) = \tfrac{1}{6}, \quad P_\Cop(00---) = \tfrac{1}{2}.
\]
These Bell probabilities are set out beside the Copenhagen probabilities in Table 2.

\begin{table}[ht]
\begin{center}
\renewcommand{\arraystretch}{1.4}
\begin{tabular}{|c|c|c|}
\hline
$abcde$ & $P_\B(abcde)$ & $P_\Cop(abcde)$  \\
\hline\hline
($00+++$) & $\tfrac{1}{3}$ & $\tfrac{1}{6}$ \\
\hline
($00++-$) & $0$ & $\tfrac{1}{6}$ \\
\hline
($00--+$) & $\tfrac{1}{6}$ & $\tfrac{1}{3}$  \\
\hline
($00---$) & $\half$ & $\tfrac{1}{3}$  \\
\hline
\end{tabular}
\end{center}
\caption{Probabilities for the histories of $F_2$ in the Frauchiger-Renner experiment}
\end{table}

The difference between the Bell and Copenhagen probabilities in Table 2 is similar to what we have already seen in Table 1 for Brukner's version of the Wigner's friend experiment. 

To be relevant to the conclusion that Frauchiger and Renner drew from their thought experiment, we must extend our concept of a sentient observer. The Frauchiger-Renner argument assumes communication between $F_1$, $W_1$ and $W_2$ and therefore requires that they should be treated as a single observer whose experience covers the measurement results of all three. The histories of this three-headed observer from $t = 0$ to $t = 5$ are of the form
\[
h(a\,a'\,a'',\ldots ,f\,f'\,f'') = \big( (a, a', a''; t = 0), \ldots, (f,f',f''; t = 5)\big)
\]
where $a,\ldots,f\in\{0,\head,\tail\},\,a'\ldots,f'\in\{0, \ok, \fail\}$ and $a'',\ldots,f''\in\{0,\ok,\fail\}$. However, the Bell and Copenhagen probabilities are both zero for any history in which there is any change in the experience of $F_1$ between $t = 1$ and $t = 3$; the relevant projection operators commute with the evolution operators $\e^{-\i H(t - 1)}$ for $ 1 < t < 3$ since all the action in this interval involves only $S$ and $F_2$. We therefore only need to consider histories 
\begin{multline*}
h(a\,a'\,a'',\; b\,b'\,b'',\; c\,c'\,c'',\; d\,d'\,d'')\\
 = \big( (a, a', a''; t = 0), (b,b',b''; t = 1), (c,c',c''; t = 4), (d,d',d''; t = 5)\big)
\end{multline*}
The two sets of probabilities for these histories are shown in Table 3, in which ``head'', ``tail'' and ``fail'' are abbreviated to their initial letters. All histories not shown in this table have $P_\B = P_\Cop = 0$.

\begin{table}[ht]
\begin{center}
\renewcommand{\arraystretch}{1.4}
\begin{tabular}{|c|c|c|}
\hline
$h$ & $P_\B(h)$ & $P_\Cop(h)$  \\
\hline\hline
$(0\,0\,0,\; \h\,0\,0,\; \h\,\f\,0,\; \h\,\f\,\f)$ & $\tfrac{3}{20}$ & $\tfrac{1}{24}$ \\
\hline
$(0\,0\,0,\; \h\,0\,0,\; \h\,\f\,0,\; \h\,\f\,\,\ok)$ & $\tfrac{1}{60}$ & $\tfrac{1}{24}$ \\
\hline
$(0\,0\,0,\; \h\,0\,0,\; \h\,\ok\,0,\; \h\,\f\,\ok)$ & $0$ & $\tfrac{1}{24}$  \\
\hline
$(0\,0\,0,\;\h\,0\,0,\;\h\,\ok\,0,\;\h\,\ok\,\ok)$ & $0$ & $\tfrac{1}{24}$  \\
\hline
$(0\,0\,0,\;\h\,0\,0,\;\t\,\f\,0,\;\t\,\f\,0$ & $\tfrac{3}{20}$ & $\tfrac{1}{24}$ \\
\hline
$(0\,0\,0.\;\h\,0\,0,\;\t\,\f\,0,\;\t\,\f\,\ok)$ & $\tfrac{1}{60}$ & $\tfrac{1}{24}$ \\
\hline
$(0\,0\,0,\;\h\,0\,0,\;\t\,\ok\,0,\;\t\,\ok\,\f)$ & $0$ &$\tfrac{1}{24}$ \\
\hline
$(0\,0\,0,\;\h\,0\,0,\;\t\,\ok\,0,\;\t\,\ok\,\ok)$ & $0$ & $\tfrac{1}{24}$ \\
\hline
$(0\,0\,0,\;\t\,0\,0,\;\h\,\f\,0,\;\h\,\f\,\f)$ & $\tfrac{9}{40}$ & $\tfrac{1}{6}$ \\
\hline
$(0\,0\,0,\;\t\,0\,0,\;\h\,\f\,0,\;\h\,\f\,\,\ok)$ & $\tfrac{1}{40}$ & $0$ \\
\hline
$(0\,0\,0,\;\t\,0\,0,\;\h\,\ok\,0,\;\h\,\ok\,\f)$ & $\tfrac{1}{24}$ & $\tfrac{1}{6}$ \\
\hline
$(0\,0\,0,\;\t\,0\,0,\;\h\,\ok\,0,\;\h\,\ok\,\ok)$ & $\tfrac{1}{24}$ & $0$ \\
\hline
$(0\,0\,0,\;\t\,0\,0,\;\t\,\f\,0,\;\t\,\f\,\f)$ & $\tfrac{9}{40}$ & $\tfrac{1}{6}$ \\
\hline
$(0\,0\,0,\;\t\,0\,0,\;\t\,\f\,0,\;\t\,\f\,\,\ok)$ & $\tfrac{1}{40}$ & $0$ \\
\hline
$(0\,0\,0,\;\t\,0\,0,\;\t\,\ok\,0,\;\t\,\ok\,\f)$ & $\tfrac{1}{24}$ & $\tfrac{1}{6}$ \\
\hline
$(0\,0\,0,\;\t\,0\,0,\;\t\,\ok\,0,\;\t\,\ok\,\ok)$ & $\tfrac{1}{24}$ & $0$ \\
\hline
\end{tabular}
\end{center}
\caption{Probabilities for the histories of $F_1$, $W_1$ and $W_2$ in the Frauchiger-Renner experiment}
\end{table}

As before, there are pairs of histories which are equally probable in the Copenhagen calculation, while Bell assigns zero probability to one of them. But now there are also histories which have $P_\Cop = 0$ but $P_\B \ne 0$, in particular those in which $F_1$ observes ``tails'' when she measures the coin at $t = 2$ and $W_1$ and $W_2$ both obtain the result ``ok'' in their measurements of $F_1$ and $F_2$. This difference is the core of the Frauchiger-Renner argument, and the substance of their claim of a problem in the description of this thought experiment by standard quantum mechanics. The zero probability in the Copenhagen calculation corresponds to $F_1$ applying the projection postulate to the universal state after seeing the result of her measurement. The non-zero probability in the Bell calculation arises if $F_1$ applies Vedral's dictum ``Unobserved results can affect future measurements'' \cite{Vedral:reality}. Knowing how the experiment is proceeding, she knows that what she observes is registered in just one component of the universal state. The whole of the universal state continues to evolve and affects the future measurements, as is explicitly included in Bell's process: the transition probabilities at any time depend on the value of the universal state at that time. The agents $F_1$, $W_1$ and $W_2$ are entitled to regard their experience as ``reality''; then in their perspective the universal state vector is not a description of reality but an influence, or force, that affects the development of reality.

We note that the history states $|K(h)\>$, as defined in \eqref{histstate}, for the histories $h$ in table 3 are not all orthogonal; in fact there are four cases in which two different histories give rise to the same history state, for example
\begin{align*}
|K(0\,0\,0,\; \h\,0\,0,\; \h\,\f\,0,\; \h\,\f\,\f)\> &= |K(0\,0\,0,\;\t\,0\,0,\;\t\,\f\,0,\;\t\,\f\,\f)\>\\
&= \frac{1}{\sqrt{6}}|\h\>_{F_1C}|\f\>_{F_2S}|\f\>_{W_1}|\f\>_{W_2}.
\end{align*}
It follows, as has been pointed out before \cite{Losada:FRhistories}, that the histories in Table 3 do not satisfy the consistency condition of \cite{Griffiths:book}. This does not mean that there is anything inconsistent in these histories; it only means that the formula \eqref{PCop} cannot consistently be applied to disjunctions of them. This can be seen, for example, by using \eqref{PCop} to calculate the Copenhagen probability of a three-time history with times $t = 0$, $t = 4$ and $t = 5$; the result is not the sum of the corresponding histories in Table 3 --- as is to be expected in standard quantum mechanics when one ignores the fact that a measurement has taken place at $t = 1$. But if the histories are considered as atoms in a probability space\footnote{Confusingly, this would normally be called an ``event space.''}, the usual (Kolmogorov) rules can be applied.

\section{Observation of History}\label{memory}

If an agent is equipped with a memory which records their momentary experience at each stage of a history, they can reduce their experience of history to a single measurement at the end of the history. Let us return to $F_2$ and suppose she has a memory with a basis $|x_1x_2x_3x_4x_5\>_M$ where $x_t$ can take values $0, +, -$ or $\times$, the final value $\times$ being recorded at times when $t$ lies in the future. Then the universal state space has an extra tensor factor $\H_M$. We denote the universal state vector in this enlarged space by $|\widetilde{\Psi}(t)\>$, with $|\Psi(t)\>$ continuing to denote the universal state in \secref{FR}. Then the state with memory is of the form
\begin{multline*}
|\widetilde{\Psi}(t)\> = \sum_{a\in\{0,+,-\}}|a\>_{F_2}|x_1x_2x_3x_4x_5\>_{M}|\psi(t)\>\\
\text{with } |\psi(t)\> \in \H_S\otimes \H_C\otimes \H_{F_1}\otimes \H_{W_1}\otimes\H_{W_2}
\end{multline*}
where $x_t = a$ and $x_s = \times$ if $s > t$. This develops as follows:
\begin{align*}
|\widetilde{\Psi}(0)\> &= |\Psi(0)\>|0\times\times\times\times\>_M,\\
|\widetilde{\Psi}(1)\> &= |\Psi(1)\>|00\times\times\times\>_M,\\
|\widetilde{\Psi}(2)\> &= \Big[ \sqrt{\tfrac{1}{6}}\big(|\fail\> - |\ok\>\big)_{F_1C}|+\>_{F_2S}|00+\times\times\>_M\\
&\quad +\sqrt{\tfrac{2}{3}}|\fail\>_{F_1C}|-\>_{F_2S}|00-\times\times\>_M \Big]|0\>_{W_1}|0\>_{W_2},\\
|\widetilde{\Psi}(3)\> &= \sqrt{\tfrac{1}{6}}|00++\times\>\Big(|\fail\>_{F_1C}|\fail\>_{W_1} - |\ok\>_{F_1C}|\ok\>_{W_1}\Big)|+\>_{F_2}|0\>_{W_2}\\
&\quad + \sqrt{\tfrac{2}{3}}|00--\times\>|\fail\>_{F_1C}|\fail\>_{W_1}|-\>_{F_2}|0\>_{W_2}\\
|\widetilde{\Psi}(4)\> &= \tfrac{1}{\sqrt{24}}|00+++\>_M\left(|\fail\>_{F_1C}|\fail\>_{W_1} - |\ok\>_{F_1C}|\ok\>_{W_1}\right)|+\>_{F_2S}\big(|\fail\> + |\ok\>\big)_{W_2}\\
& + \tfrac{1}{\sqrt{24}}|00++-\>_M\left(|\fail\>_{F_1C}|\fail\>_{W_1} - |\ok\>_{F_1C}|\ok\>_{W_1}\right)|+\>_{F_2S}\big(|\fail\> - |\ok\>\big)_{W_2}\\
& + \tfrac{1}{\sqrt{6}}|00--+\>_M|\fail\>_{F_1C}|\fail\>_{W_1}|+\>_{F_2S}\big(|\fail\> -|\ok\>\big)_{W_2}\\ 
&+ \tfrac{1}{\sqrt{6}}|00---\>_M|\fail\>_{F_1C}|\fail\>_{W_1}|-\>_{F_2S}\big(|\fail\> +|\ok\>\big)_{W_2} .
\end{align*}
From this we can read off the probabilities of the memory readings as
\[
P(00+++) = P(00++-) = \tfrac{1}{6}, \quad P(00--+) = P(00---) = \tfrac{1}{3}, 
\]
the same as the Copenhagen probabilities in Table 1. The reason for this is clear: the entanglement of the original system with the memory serves to separate the sensible states of $F_2$, so that the time development of each such state proceeds as if the other states had been annihilated by a projection operator. This prevents subsequent interference between these components: entanglement implements collapse.

This calculation illustrates a general fact: if a system is linked to a memory which keeps a permanent record of a set $B$ of basis states of the system, then the probabilities which will be observed in the memory are the same as those which would be calculated in a Copenhagen calculation assuming that the system (without the memory) undergoes collapse at each time stage onto a state of the basis $B$. Thus a direct attempt to establish the probabilities of various histories of the system will lead to the Copenhagen probabilities. This might be seen as a justification of the probabilities assumed in the ``consistent histories'' version of quantum mechanics. However, this does not rule out the possibility of indirect effects of Bell probabilities if the system is not coupled to a memory, in a similar way to the quantum effects of superposition which are not visible if the system undergoes measurement. Moreover, this justification of Copenhagen probabilities is only valid if the system whose histories are in question is not the whole universe; it depends on there being something (the memory) external to the system. This undercuts the claim of consistent-histories theory to be a version of quantum mechanics which is specially adapted to cosmology.

\section{Conclusion}

We have explored an alternative theory of histories --- sequences of events, such as measurements --- in quantum mechanics. These have usually been treated by assuming the Dirac/von Neumann/L\"uders projection postulate of the Copenhagen formulation of quantum mechanics. We have shown that this is not necessary; there is an alternative formulation in which this postulate is not assumed, but histories can still be treated. We have shown that this gives different results for the probability of a history. However, this difference is not directly observable: a simple empirical procedure for observing histories is bound to give the same results as if the projection postulate was valid. Nevertheless, this means that any theory that assumes this postulate, such as the ``consistent histories'' interpretation of quantum mechanics, cannot be a theory of the whole universe and cannot be a basis for cosmology.

\linespace

\noindent {\bf Acknowledgement}

I am grateful to Flavio del Santo and his co-workers for a valuable discussion, and to an anonymous referee for suggestions which have greatly improved the presentation of these ideas.

%\bibliography{quantum}
%\bibliographystyle{plain}

\end{document}